\documentclass{elsarticle}
\usepackage{amssymb}
\usepackage{pifont}
\usepackage{array}
\usepackage{tabularx}
\usepackage{booktabs}
\usepackage{hyperref}
\usepackage[table]{xcolor}
\usepackage{rotating}

\journal{Information and Software Technology}
\graphicspath{ {./} }
\begin{document}
\begin{frontmatter}
    \title{Resolvi: A Reference Architecture for Extensible, Scalable and Interoperable Entity Resolution}
    \author[1]{Andrei Olar}
    \ead{andrei.olar@ubbcluj.ro}
    \affiliation[1]{
        organization={Mathematics and Computer Science Faculty, Babes-Bolyai University},
        addressline={Str. M. Kogalniceanu, nr. 1},
        postcode={400084},
        postcodesep={},
        city={Cluj-Napoca},
        country={Romania}
    }

    \begin{abstract}
        \textbf{Context:} Entity resolution (ER) plays a pivotal role in data
        management by determining whether multiple records correspond to the
        same real-world entity.
        Because of its critical importance across domains such as healthcare,
        finance, and machine learning and its long research history designing
        and implementing ER systems remains challenging in practice due to the
        wide array of methodologies and tools available.
        This diversity results in a paradox of choice for practitioners, which
        is further compounded by the various ER variants~(record linkage, entity
        alignment, merge/purge, a.s.o).

        \noindent\textbf{Objective:} This paper introduces \textit{Resolvi}, a
        reference architecture for facilitating the design of ER systems.
        The goal is to facilitate creating extensible, interoperable and
        scalable ER systems and to reduce architectural decision-making duration.

        \noindent\textbf{Methods:} Software design techniques such as the 4+1
        view model or visual communication tools such as UML are used to present
        the reference architecture in a structured way.
        Source code analysis and literature review are used to derive the main
        elements of the reference architecture.

        \noindent\textbf{Results:} This paper identifies generic requirements
        and architectural qualities of ER systems.
        It provides design guidelines, patterns, and recommendations for
        creating extensible, scalable, and interoperable ER systems.
        Furthermore, it highlights implementation best practices and deployment
        strategies based on insights from existing systems.

        \noindent\textbf{Conclusion:} The proposed reference architecture offers
        a foundational blueprint for researchers and practitioners in developing
        extensible, interoperable, and scalable ER systems.
        \emph{Resolvi} provides clear abstractions and design recommendations
        which simplify architecture decision making, whether designing new ER
        systems or improving existing designs.
    \end{abstract}
\end{frontmatter}

\section{Introduction}

Entity resolution (abbrev. ER) is the task of determining whether
multiple information records refer to the same real-world entity.
Despite the straightforward definition, ER is supported by a vast field of
research and has critical applications in domains such as
healthcare~\cite{newcombe1959,medicalrecordlinkage2012}, finance~\cite{finnerd2022},
population censuses~\cite{censusrecordlinkage2011}, and fraud detection~\cite{frauder2019}.
It also plays a fundamental role in data integration, data cleaning, and machine
learning pipeline preparation~\cite{christen2012data,e2e2020,4generationser2021}.

The terminology surrounding ER is diverse.
It is commonly referred to as record linkage, deduplication, merge/purge, or
entity matching~\cite{e2e2020,4generationser2021,almostall2022}.
In some contexts, it appears under the names entity alignment or named entity
recognition and disambiguation (NERD)~\cite{nerd2012}.
This terminological variation reflects the breadth and complexity of the field,
creating challenges in unifying methodologies and system architectures.

Although ER has been extensively studied, existing literature primarily focuses
on individual solutions, with generic systems being far less common.
While existing surveys~\cite{e2e2020,4generationser2021,almostall2022} provide
valuable overviews, they often contribute to the paradox of
choice~\cite{schwartz2015}, as practitioners must navigate an overwhelming
number of methodologies, tools, and frameworks.
Each generic system promotes its own original design, contributing to choice
overload.
Notably, no reference architecture exists that systematically captures common
design principles and implementation patterns for ER systems.
The absence of such an architecture hinders standardization across ER
implementations which, in turn, raises ER adoption costs.

This \textbf{paper introduces an ER reference architecture} that synthesizes
insights from existing ER systems and literature to provide a structured
foundation for ER system design.
The research is guided by two primary questions:

\textbf{
    RQ1: What are the generic use cases of entity resolution systems, and how do
    these use cases inform the design of abstract components and processes
    commonly found in entity resolution architectures and implementations?
}

\textbf{
    RQ2: Does the entity resolution problem admit a reference architecture, or
    are there boundaries past which generalization and abstraction lose their
    usefulness?
}

The rest of the paper is organized as follows.
Initially, the focus is on providing background information on related
work and the reference architecture concept, quickly moving to establishing a
standard vocabulary for ER concepts within the reference architecture and
highlighting reference ER system implementations.
Then, the narrative naturally veers towards generic requirements, use cases and
logical components.
The centerpiece of the paper is the attempt to describe the ER process at
runtime in an abstract way, encompassing as many ER flavours as possible.
This is followed by a discussion around design and implementation guidelines,
deployment considerations, the feedback process for the reference architecture
and the above research questions before concluding.

\section{Background}\label{sec-background}

\subsection{Related Work}

The idea to study generic architectural concepts in the sphere of ER was kindled
by reading excellent reviews~\cite{e2e2020,almostall2022,eakit2021,openea2020,liu2023,deepmatcher2018}
on the subject.
These surveys provided valuable overviews of ER and an entry point to discover
ER systems.
The reference architecture itself was crafted by inspecting the source code of
deduplication or record linkage systems~\cite{recordlinkage2019,febrl2008,silk2013,dblink2021,oyster2013,deepmatcher2018,duke2014,ditto2020,fastlink2019,splink2022,zingg2024,dedupe2022,famer2018},
while entity alignment concepts were gathered from systems which are based on
literature reviews~\cite{openea2020,eakit2021}.

In the absence of a comprehensive review focusing on the disambiguation aspect
of named entity recognition and disambiguation, several independent sources were
studied to better understand the problem.
The opinion that named entity disambiguation systems are designed very similarly
to merge/purge systems is rooted in the NERD framework~\cite{nerd2012} itself,
which implements a special ontology-based merge algorithm of extracted named
entities.
FOX~\cite{multilingualfox2017} uses building blocks which are prevalent in
record linkage systems, too~(e.g indexing components, external knowledge bases).
Modern systems such as Ensemble Nerd~\cite{ensemblenerd2018} use separate black
box subsystems (in this case neural networks) to perform feature extraction and
matching, similarly to DeepMatcher~\cite{deepmatcher2018}.
Finally, to the best of our knowledge there is no dispute over the need to
extract structured information from unstructured text across the named entity
recognition literature.

To construct a viable reference architecture, examples of existing reference
architectures~\cite{refarchbigdata2015,comprefarch2024} as well as reference
architecture frameworks~\cite{refarchframework2012,refarchconsolidate2014} and
studies~\cite{refarch2010,refarch3decades2021} were consulted.
Ideas such as documenting patterns and joining business and technical contexts
in a reference architecture were inspired by the Rational Unified Process
(RUP)~\cite{rup2004}.
The existence of multiple perspectives over the same rough generic concepts was
inspired by the 4+1 View Model~\cite{4+1krutchen1995}, also used in RUP.\@
Lastly, UML 2.5~\cite{uml2.5_2017} was used in most figures in order to adhere
to a standard visual communication framework.

\subsection{The Reference Architecture Concept}

We begin by defining what constitutes a reference architecture. One perspective
describes it as \emph{
    capturing the essence of existing architectures, and the vision of future
    needs and evolution to provide guidance to assist in developing new system
    architectures
}~\cite{refarch2010}.
Another definition emphasizes \emph{
    an abstraction of software elements, together with the main responsibilities
    and interactions of such elements, capturing the essentials of existing
    software systems in a domain and serving as a guide for the architectural
    design of new software systems (or versions of them) in the domain
}~\cite{refarch3decades2021}.
Both definitions highlight the dual purpose of reference architectures:
synthesizing existing knowledge and offering structured guidance for future
developments.

In practice, reference architectures consolidate architectural knowledge and
best practices, ensuring systematic reuse and providing a standardized blueprint
for system design~\cite{refarch2010,refarch3decades2021}.
Beyond their conceptual role, they offer concrete benefits, particularly in
standardization, where they establish common terminology, models, structures
and processes.
Existing systems serve as examples, illustrating how abstract architectural
concepts manifest in practice.

To enhance clarity and consistency, a standard design language is usually
adopted to represent architectural components.
In crafting this reference architecture the 4+1 view model~\cite{4+1krutchen1995}
was adapted to provide multiple points of view over abstract concepts.
Relevant design patterns~\cite{gofdesign1994} to the target domain and
implementation best practices were identified.

Reference architectures evolve through structured methodologies, incorporating
empirical insights based on collaboration to enhance maturity and
adoption~\cite{refarch3decades2021}.
Several frameworks (e.g.,~\cite{refarchframework2012,refarchconsolidate2014})
exist to support the development of reference architectures, providing
methodological guidance.
Their usefulness to the reader lies in how they categorize reference
architectures.
For instance, the current reference architecture is a ``Type 3'' architecture
according to~\cite{refarchframework2012}.
It was derived from observations made on existing ER systems and may be used
only to facilitate creating new system architectures or to align existing
architectures.

\subsection{ER Vocabulary}\label{subsec-background-vocabulary}

One of the important functions of a reference architecture is to establish
common ground through the standardization of terminology in the target domain.
The terminology used in entity resolution literature is often confusing.
For instance, papers by the same authors use different terminology to describe
the same concept (e.g ``description''~\cite{e2e2020} and
``entity profile''~\cite{blocking2020}).
Table~\ref{tbl-er-terminology} provides a definition list containing the terms
used in this reference architecture as they relate to the ER problem.

\begin{table}[h!tp]
    \caption{
        Entity resolution terminology.
    }
    \begin{center}
        \begin{tabular}{ll}
            Term               & Definition                                  \\
            \toprule
            data source        & a source of information which contains      \\
                               & data that potentially refer to a real-world \\
                               & object                                      \\
            information record & a discrete chunk of information from a      \\
                               & data source                                 \\
            record             & \textit{information record}                 \\
            data record        & \textit{information record}                 \\
            extraction trait   & data extraction algorithm that can process  \\
                               & information records in a relevant way for   \\
                               & entity resolution                           \\
            entity reference   & potential reference to a real-world object  \\
                               & (typically obtained from a single data      \\
                               & source using a selection of extraction      \\
                               & traits)                                     \\
            entity profile     & representation of a real-world object       \\
                               & based on information from the available     \\
                               & data sources (typically obtained from       \\
                               & multiple entity references)                 \\
            entity             & a real-world object                         \\
        \end{tabular}
    \end{center}\label{tbl-er-terminology}
\end{table}%

The reference architecture describes two aspects of ER systems.
More commonly, ER is a process which outputs entity profiles based on entity
references from input data sources.
This process is referred to as the ER \textbf{\emph{runtime}}.
A second equally common aspect of ER systems involves configuring the ER
runtime and evaluating the performance of ER.\@
This constitutes the \textbf{\emph{design time}} aspect of an entity resolution
system.

\subsection{Reference ER Systems}\label{subsec-background-reference-systems}

As already established, ER has many flavours with names that reflect specific
modi operandi.
Some flavours can often be melded into a single, more dominant flavour.
For instance, deduplication can be viewed as an instantiation of merge/purge.
In turn, the merge/purge problem can be viewed a special case of record linkage
where the record linkage process mandatorily ends with the specific action of
merging linked entity references in-place in entity profiles instead of allowing
an open course of actions for dealing with the linked entity references.
Lastly, even though it is sometimes described as a mere step in the ER
runtime~\cite{oyster2013,jedai2018,dedupe2022}, we consider ``entity matching''
to be a particular ER flavour (record linkage flavour, to be exact) where the
final entity profiles are not obtained as a result of clustering linked entity
references.

Having this in mind, the reference architecture focuses on record linkage
systems because of its seeming centrality to the ER problem.
It is also the most studied ER flavour, with the longest history, going back
more than half a century~\cite{newcombe1959,fs1969}.

Other ER flavours follow the outline of the record linkage runtime process, too.
Even though the implementation details are different, entity alignment or NERD
systems can be implemented using similar design concepts.
The departure from record linkage seems to be the most pronounced in the case of
entity alignment which operates on knowledge graphs~\cite{openea2020,eakit2021}.
This reference architecture will do its best to supplant the divergences
specific to entity alignment with suggestions on how to adhere to its generic
abstract outlook.
On the other hand, most of the abstractions described in the reference
architecture should be applicable to designing NERD systems with only minor
changes.

The following principle guides the selection process for reference ER systems:
``proven architectures of past and existing products are transformed in a
reference architecture''~\cite{refarch2010}.
To increase the reference architecture's verifiability, only systems which meet
the following criteria were taken into account:

\begin{itemize}
    \item code inspection availability;
    \item end-to-end (E2E) entity resolution implementation;
    \item actively maintained.
\end{itemize}

The large number of candidate ER systems remaining applying the above criteria
prompted filtering the remaining systems based on their implementation of
key aspects were of particular subjective interest.
The final selection is available in Table~\ref{tbl-er-systems}.

\begin{table}[htp]
    \caption{
        Systems that implement entity resolution.
        Each system was chosen for highlighting a key aspect.
    }
    \begin{center}
        \begin{tabularx}{\textwidth}{lcl}
            Name                                                           & Technology & Key Aspect                        \\
            \toprule
            \texttt{ditto}~\cite{ditto2020} + Magellan~\cite{magellan2018} & Python     & Pre-trained Neural Networks       \\
            \texttt{deepmatcher}~\cite{deepmatcher2018}                    & Python     & Neural Network Architecture       \\
            \href{https://github.com/larsga/Duke}{Duke}                    & JVM        & Automated Rule Configuration      \\
            JedAI Toolkit~\cite{jedai2018}                                 & JVM        & Core Pipeline Architecture        \\
            FAMER~\cite{famer2018}                                         & JVM        & Distributed Graph Linkage         \\
                                                                           &            & Incremental ER                    \\
            Splink~\cite{splink2022}                                       & JVM        & Distributed Probabilistic Linkage \\
            d-blink~\cite{dblink2021}                                      & Python     & Probabilistic Clustering          \\
        \end{tabularx}
    \end{center}\label{tbl-er-systems}
\end{table}%

Systems that were not open source~\cite{fusion2020} at the time of writing and
ones that did not seem to be actively maintained~\cite{febrl2008} were also
evaluated via peer reviewed articles published about them.
Other systems stood out, but were architecturally very similar to one or more
of the reference ER systems.
This includes systems like the Silk Framework~\cite{silk2013}, Zingg~\cite{zingg2024},
OYSTER~\cite{oyster2013}, FastLink~\cite{fastlink2019}, Dedupe~\cite{dedupe2022}
or RecordLinkage~\cite{recordlinkage2019} all of which have a runtime design
which is very similar to the JedAI Toolkit~\cite{jedai2018}.

Our review of the existing ER systems yielded some preliminary observations.
Active learning solutions seemed to be prevalent at design
time~\cite{dedupe2022,zingg2024}.
Most of the systems we surveyed were able to provide some means to evaluate ER
quality.
Our subjective preference for ER systems which automate design time choices
promoted Duke~\cite{duke2014} to our list of reference systems due to its
genetic algorithm which determines optimal record linkage configurations without
user participation.

Performance and scalability are common cross-cutting concerns brought up in the
ER space~\cite{4generationser2021}.
This lead to promoting d-blink~\cite{dblink2021},
Splink~\cite{splink2022} and FAMER~\cite{famer2018} as reference systems because
they were designed on top of generic runtimes designed for distributed
computing.

Most of the extensibility built into existing systems focuses on the key aspects
of ER (matching, clustering, data preprocessing).
The Magellan~\cite{magellan2018} framework for entity matching is extensible in
a completely different way: it allows using matching components such as
Ditto~\cite{ditto2020} or replacing the entire rule-based ER runtime with a full
neural network architecture designed for ER such as DeepMatcher~\cite{deepmatcher2018}.
This emphasizes the higher extensibility of ER systems with a framework design.

\section{ER Requirements}

The architecture of software systems is linked to both the business context and
the technical context~\cite{refarch2010,refarchframework2012} of the problem
they attempt to solve.
The requirements related to the business context constitute the
\textit{problem domain}~\cite{fosa2020}.
The technical context is typically composed from the system requirements which
define the system's capabilities, also known as
\textit{architectural characteristics}~\cite{fosa2020}.

\subsection{Business Requirements}

Part of the reason for creating a reference architecture is to identify goals
that all system architectures implementing the reference architecture will
adhere to.
In the case of entity resolution, there are requirements which determine whether
a system performs entity resolution or some other task.

The first such requirement is determining whether multiple information refer to
the same \textit{real world object}.
With the risk of stating the obvious, this entails that the system should
discern what a real world object is.
Human assistance in this regard is arguably required at this point in our
technological development.
It logically follows that completely unsupervised entity resolution solutions
are more unreliable than supervised ones.
Consequently, \emph{humans should at least be allowed in the loop at design time}.

The second requirement that results directly from the definition of ER is to
return \textit{all} references to the same real world object from the presented
input data.
This requirement entails that the presented output is necessarily either a
single entity reference or the result of a merge operation between multiple
references.
In this context, the variability of the output over time in contrast with the
stability of the resolved real-world object is highlighted~\cite{fusion2020}.
Taking the idea a step further, the \emph{entity profiles resolved today might
    be used as entity references tomorrow}.
Allowing extension in this regard could be beneficial.

Users might want to inspect all entity profiles resulting from a single scan of
multiple data sources.
This is known as \textbf{batch ER}~\cite{e2e2020}.
The alternative is to feed the ER system information one record at a time.
Entity profiles which refer to the same entity are retrieved progressively
with each input information record.
This is known as \textbf{incremental ER}~\cite{incrrl2014}.
\emph{ER systems may implement one or both variants}.

\subsection{Architectural Characteristics}\label{subsec-erreq-arch-characteristics}

Just as business requirements dictate whether a system performs entity
resolution, certain architectural characteristics are universally required for
systems that execute end-to-end entity resolution.

ER is a highly polymorphic task with a huge number of applications.
Among the top priorities when implementing any such system is its
\textbf{extensibility}.
Extensibility is exhibited to various degrees in practice today.
Perhaps the most important reason for making extensibility the top priority of
this reference architecture is the variability of implementation choices made
by the existing ER systems.
Whether discussing a customizable number of data sources~\cite{famer2018,jedai2018}
or the ability to completely replace the core runtime logic~\cite{magellan2018,deepmatcher2018},
ER systems today are nothing if not flexible.
The technological diversity ranging from systems using SQL~\cite{splink2022} to
ones using advanced neural networks~\cite{ditto2020} is another reason for the
reference architecture to prioritize extensibility.
This prioritization necessarily entails a higher degree of generalization and
abstraction.

ER is valuable standalone or as part of a more complex process.
Therefore, the best ER systems will be highly interoperable either by supporting
many data contracts or by using widely adopted technologies and data
formats.
\textbf{Interoperability} is an obvious concern to some of the inspected
reference systems~\cite{jedai2018,duke2014,famer2018,splink2022}.
Because the reference architecture values interoperability, the concepts and
recommendations presented favour the adoption of frameworks or standards
wherever possible.

Basic ER implementations that rely on similarities between entity references
have a minimum time complexity of $\mathcal{O}(n^2)$~\cite{e2e2020} where $n$ is
the total number of entity references available to the system.
The high complexity coupled with the need to process large amounts of data makes
performance and scalability indispensable architectural characteristics of any
modern ER system.
\textbf{Performance} is the ability of a system to use the available hardware
resources efficiently~\cite{fosa2020} (i.e the system does not waste
computations).
\textbf{Scalability} is the ability of a system to use hardware resources
effectively as the demand on the system increases~\cite{fosa2020} (e.g the
system is able to access more resources as it receives more input).
As a consequence of valuing performance and scalability, the reference
architecture insists on aspects which promote these qualities, such as blocking,
filtering, parallel execution or distributed computation.

\section{ER Use Cases}

It is important to distinguish the problem domain of specific ER systems from
generic use cases which apply to all ER systems.
This section goes over three collections of use cases that are present in all of
the reference ER systems.
Each collection is represented with a use case diagram.
Use case diagrams are a type of UML diagram used to represent the dynamic
aspect~\cite{uml1998} of a system.

\subsection{Roles}

Use cases are linked to actors who fulfill certain roles.
\textbf{End users} are regular users who want to run ER jobs to assemble entity
profiles.
\textbf{Administrators} are members of staff who set up the ER system and have a
basic understanding of ER.\@
This role may correspond to real-world roles such as ``data engineer'', ``system
administrator'' or ``data scientist''.
\textbf{Developers} create new functionality that is connected to ER.\@
They may assist administrators in setting up the functionality that they
developed.
This role may correspond to real-world ``software engineer'', ``data engineer'',
``data scientist'' or ``machine learning engineer'' roles.

The generic use cases are presented under the assumption of a multi-user
organisation.
This choice is justified by the high costs of running ER and the need to involve
multiple people in the setup process.
Note that there is no assumption about authentication or, indeed, security
in general because these concerns vary with the problem domain.

\subsection{Entity Resolution Runtime}\label{subsec-uc-runtime}

\begin{figure}[htbp]
    \begin{center}
        \includegraphics[width=\textwidth]{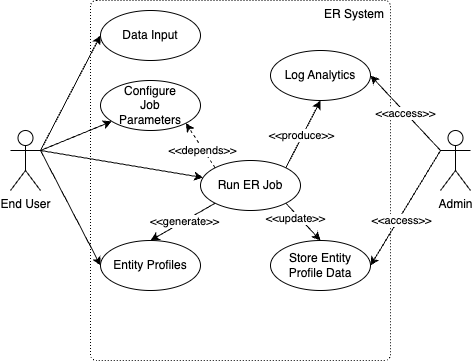}
        \caption{
            At runtime several use cases of the ER system are accessed by end
            users and admins.
            The ``runtime'' is detailed in Subsection~\ref{subsec-uc-runtime}.
        }\label{fig-uc-runtime}
    \end{center}
\end{figure}

Figure~\ref{fig-uc-runtime} shows the collection of use cases associated with
the ER \textit{runtime}.
All existing systems implement core ER use cases such as supplying the system
with input data, configuring parameters, running ER and allowing inspection
of entity profiles.

Even though existing systems prioritized enabling zero-configuration entity
resolution, the option to run ER irrespective of the dataset is not yet
available.
Currently the best choice is between a careful selection of dataset features
relevant for entity resolution~\cite{splink2022} or using active
learning~\cite{lowrestransactiver2019,transer2022}.
Genetic programming~\cite{duke2014} offers additional hope for reducing user
involvement in ER process parametrization.

Some ER systems store entity references and entity profiles for later use.
Storing the state is particularly important for performing progressive
ER~\cite{famer2018,jedai2018,duke2014} or to preserve stable identifiers across
time for entity profiles~\cite{fusion2020}.

System architectures should center the design of their runtime around end users.
Additional system use cases such as storing a runtime state should be designed
with administrators in mind, too.
It is strongly recommended that systems architectures focus on performance
monitoring as well as other analysis and observability mechanisms which
facilitate understanding the ER runtime.
These mechanisms are particularly important to administrators.

\subsection{System Configuration}\label{subsec-uc-config}

\begin{figure}[htbp]
    \begin{center}
        \includegraphics[width=\textwidth]{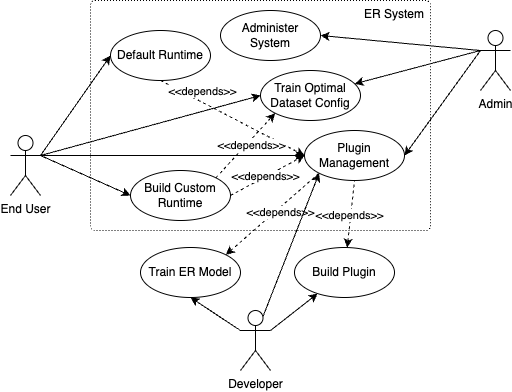}
        \caption{
            ER system configuration from Subsection~\ref{subsec-uc-config}.
            Building custom runtimes and choosing a default runtime are
            interesting to the end user.
            Administering the system and its plugins falls unto the
            administration personnel.
            Developing plugins and training ER models is a task for developers.
        }\label{fig-uc-config}
    \end{center}
\end{figure}

Configuration is performed on three levels in an ER system: ER runtime
parameters, runtime customization and system configuration.
The first level which addresses parameterizing the ER runtime has already been
described in Subsection~\ref{subsec-uc-runtime}.
Figure~\ref{fig-uc-config} shows the use cases connected to the other two levels.

In existing systems, the end user can customize the runtime itself by choosing a
default runtime~\cite{splink2022,famer2018} or even building one from existing
components exposed by the system~\cite{silk2013,jedai2018}.
Another runtime customization is the ability to train a configuration optimised
for a specific dataset either via active learning~\cite{zingg2024} or genetic
algorithms~\cite{duke2014}.
This configuration can then be attached to many ER runtime executions.
This approach stands in contrast to building the optimal configuration at
runtime through unsupervised learning~\cite{splink2022}.
In that case a runtime configuration of the data record attributes that are
relevant in ER is sufficient.

System configuration is exhibited in existing ER systems through plugin
management for some~(DeepMatcher~\cite{deepmatcher2018} can use
Magellan~\cite{magellan2018} for entity reference extraction) or for all of the
ER runtime components~(both FAMER~\cite{famer2018} and JedAI~\cite{jedai2018}
allow extending almost all components used at runtime).
Some systems are designed as applications to be used by end users.
JedAI~\cite{oyster2013} and FAMER~\cite{famer2018} are in this category.
These systems typically implement a runtime paradigm~((rule-based or
graph-based, respectively)) which cannot be changed through configuration.
In contrast, systems may be designed as frameworks~\cite{silk2013,magellan2018},
enabling the configuration of the runtime architecture itself.
The DeepMatcher design~\cite{deepmatcher2018} allows using neural networks
throughout the entire runtime or only a part of it by allowing the ER
application developers to replace the way embeddings are generated, for example.
This reference architecture favours framework design because it is more
extensible.
ER system designs should allow swapping out some or all runtime components at
design time without changing the underlying abstractions presented within this
reference architecture.

Specific ER system architectures should involve \textit{administrators} in most
configuration use cases.
This recommendation assumes that administrators are better prepared on security
and configuration topics than end users and that in practice they usually have
broader access to the system than developers.
\textit{Developers} also play a role in the design of the ER runtime.
They may train machine learning models or implement plugins that perform various
ER tasks such as matching, clustering or data ingestion.
These models and plugins are visible to end users through configuration.
Developers may register and unregister the models and plugins they author.
They also typically maintain plugin implementations and models.

Plugin management may be performed by all three roles: administrators, end
users and developers.
End users might focus on choosing the plugins they need to configure custom
runtimes, whereas developers would focus on managing their own plugins.
Finally, administrators could enforce system-wide rules for using plugins.
A system architecture that allows external developer involvement (through plugin
management or otherwise), may be desirable.
However, it raises many potential concerns, primarily from the sphere of data
privacy and security.
Careful consideration must be offered when designing solutions with this level
of extensibility.

We saw that configuration can occur at three levels: parametrization, runtime
configuration, and system configuration.
The configuration of an ER system depends on its technical goal, problem domain,
and technology context.
This limits the reference architecture, as no universal recommendations or
abstractions apply solely to ER systems.
The best approach is to follow established design principles and best practices
when defining logical components and the configuration processes.
Throughout the remainder of this paper we will not insist on the configuration
of ER systems.

\subsection{Entity Resolution Evaluation}\label{subsec-uc-eval}

\begin{figure}[htbp]
    \begin{center}
        \includegraphics[width=\textwidth]{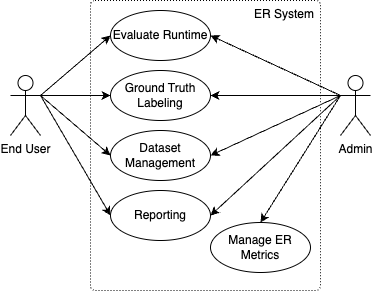}
        \caption{
            ER system configuration from Subsection~\ref{subsec-uc-eval}.
            Building custom runtimes and choosing a default runtime are
            interesting to the end user.
            Administering the system and its plugins falls unto the
            administration personnel.
            Developing plugins and training ER models is a task for developers.
        }\label{fig-uc-eval}
    \end{center}
\end{figure}

ER evaluation is described in literature in terms of both the
performance~\cite{schemaagnosticproger2019,blocking2020} and the
quality~\cite{algebraicermetrics2007,ereval2010,menestrinaermetrics2010,maidasaniermetrics2012}
of ER.\@
Additionally, standalone libraries for computing ER metrics~\cite{erevaluation2022,pyresolvemetrics2024}
and systems for comparing evaluation reports~\cite{fever2009} were developed.
Although ER evaluation methods rely on ground truth information supplied by end
users or administrators, there are approaches which discuss evaluation
statistics that work even in the absence of a benchmark dataset~\cite{clusterstats2024}.
This interest is driven by the desire to reduce the labeling effort which
accompanies the creation of a relevant evaluation dataset~\cite{seqsampling2017,clusterstats2024}.
Figure~\ref{fig-uc-eval} shows the use cases surrounding the evaluation of ER.\@

Existing systems allow end users to label comparison results as true or false
positives (known as ground truth labeling) in active learning
scenarios~\cite{dedupe2022,zingg2024}.
Other systems~\cite{oyster2013,jedai2018} use the same labeling mechanism to
create and manage benchmark datasets for specific problem domains.
All reference systems allow evaluating the ER runtime, and consulting the
evaluation results in reports.
Benchmark datasets are usually provided by the system, but the ability to manage
them varies even between the reference systems with some~\cite{jedai2018} being
more proficient than others~\cite{duke2014}.

In our view, end users would use the ER system to evaluate the ER runtime, label
ground truth data, manage benchmark datasets or view ER evaluation reports.
Administrators would use the system similarly to users, with an added ability
to configure the available metrics for evaluating the ER runtime.

System architectures derived from this reference should incorporate ER
evaluation use cases.
In terms of a recommended logical or procedural approach, there is none.
The reason behind this is that ER systems may be evaluated on design time or
at runtime.
Depending on the above choice, evaluation can be very simple~(similar to what
all existing systems implement today) or very complex~(real-time monitoring of
the results, for example).
While ER evaluation design is important, it is adjacent to ER runtime design.
Furthermore, reference implementations of ER evaluation systems which focus on
the real-time evaluation of ER at runtime are missing, to the best of our
knowledge.
Throughout the remainder of this paper we will not dive deeper into topics
related to ER evaluation, leaving that expansion on this paper as future work.

\section{ER Logical Components}~\label{sec-lv}

This section focuses on highly abstract and generic logical components of the
ER runtime.
The configuration and evaluation use cases can be designed by using generic
principles and, because of the multitude of choices, there do not seem to be
standard abstractions or design patterns which apply to the ER problem in ways
that should be highlighted.

Reference architectures are used to evidentiate design patterns~\cite{refarchframework2012}
and the \textbf{natural structuring of the ER runtime as a pipeline} is a key
observation in this regard.
All reference ER systems exhibit pipeline structures even though the specific
details of each implementation vary widely.
The main components of the ER pipeline exhibited in the reference systems'
designs correspond to existing literature on the subject~\cite{e2e2020}: input
data processor, comparison preparation, matching engine, clustering engine, and
entity profile presentation layer.

The pipeline design pattern is a well-established concept in both literature and
practice~\cite{pipeline1995, rethinkingpipeline2004}.
It structures computation as a sequence of processing components, where each
component receives input, transforms it, and passes the output to the next stage
in the pipeline. This design pattern promotes modularity, configurability, and
scalability, allowing each component to operate independently and process data
as soon as it becomes available~\cite{pipeline1995}.
The pipeline structure closely mirrors neural network architectures, where
layers sequentially process and refine representations.
Prior work has successfully leveraged this structure for neural network-based ER
systems~\cite{deepmatcher2018}, since ER similarly involves sequential
transformations of data.

Unlike the traditional pipeline design, which enforces a rigid sequence of
execution~\cite{pipeline1995}, the existing reference systems exhibit a more
flexible version of the pipeline design, closer in kind to the State-based
pipeline model~\cite{rethinkingpipeline2004}.
System architectures may determine which components of the pipeline are
applicable to their specific problem domain and system constraints.
This flexibility supports various pipeline types—classic, neural network-based,
progressive, or hybrid—allowing each implementation to define its own components
and execution logic.

\subsection{Information Flow}\label{subsec-information-flow}

\begin{figure}[htbp]
    \begin{center}
        \includegraphics[width=\textwidth]{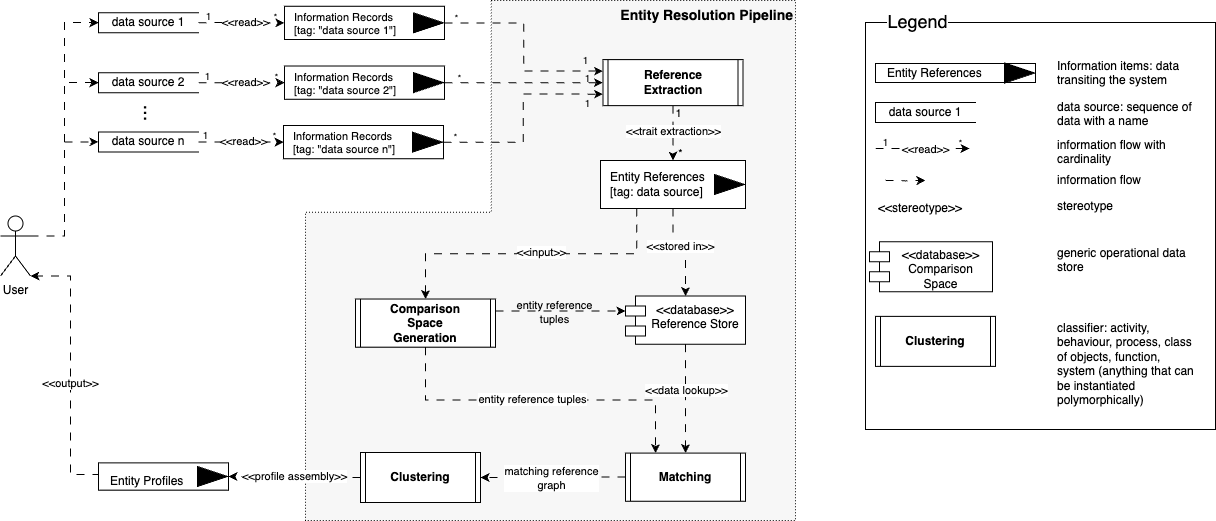}
        \caption{
            Batch entity resolution information flow at runtime, as described in
            Subsection~\ref{subsec-information-flow}.
        }\label{fig-lv-batch-er}
    \end{center}
\end{figure}

The pipeline nature of entity resolution (ER) becomes evident when analysing its
information flow at an abstract level.
Figure~\ref{fig-lv-batch-er} illustrates the runtime information flow for that
which is termed \emph{batch ER} in literature~\cite{e2e2020}.

In batch ER, input data consists of multiple data sources containing information
records, typically provided by the end user.
The representation of these records varies by source type and often requires
dedicated adapters, as seen in existing implementations~\cite{jedai2018,duke2014,silk2013,famer2018}.
Some systems support streaming data sources~\cite{famer2018}, effectively
removing the upper bound on the number of processable records.

This lack of an upper bound can lead to performing what is known as
\emph{incremental ER}~\cite{incrrl2014,e2e2020}.
It differs from batch ER in two key aspects: the requirement for maintaining a
runtime state and the continuous ingestion of input data rather than batch
processing.
Continuous ingestion also introduces the possibility of supporting certain
architecture styles better~(e.g service oriented architecture, event driven
architecture).

\begin{figure}[htbp]
    \begin{center}
        \includegraphics[width=\textwidth]{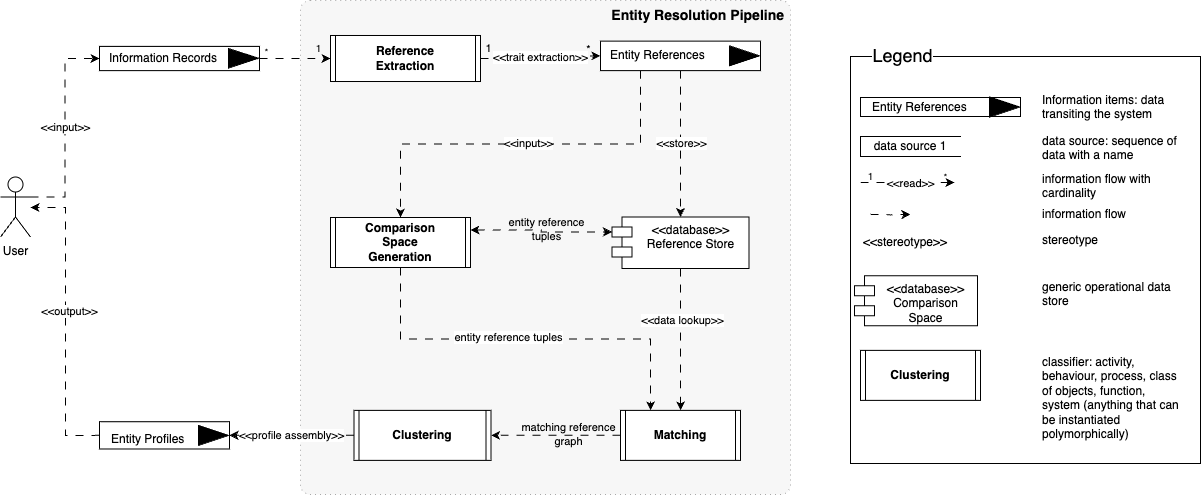}
        \caption{
            Incremental ER information flow at runtime, as described in
            Subsection~\ref{subsec-information-flow}.
        }\label{fig-lv-incremental-er}
    \end{center}
\end{figure}

Due to these runtime differences, incremental ER systems are typically
implemented as long-lived processes, whereas batch ER solutions are commonly
discrete jobs.
However, this distinction does not preclude leveraging established distributed
processing frameworks, such as Apache Spark~\cite{splink2022} or Flink~\cite{famer2018},
to manage long-running processes capable of handling multiple parallel requests.
In the context of incremental ER, Figure~\ref{fig-lv-incremental-er} highlights
the differences at the individual ER job level, assuming a long-running process
designed to handle concurrent requests.

\begin{table}[htp]
    \begin{center}
        \begin{tabularx}{\textwidth}{lccl}
            System                                                         & Incremental & Batch      & Source Code                                                                                                                                                                                                                      \\
            \toprule
            \texttt{ditto}~\cite{ditto2020} + Magellan~\cite{magellan2018} & \ding{56}   & \checkmark & \href{https://raw.githubusercontent.com/megagonlabs/ditto/refs/heads/master/run\_all\_er\_magellan.py}{\texttt{\scriptsize{ditto.run\_all\_er\_magellan}}}                                                                       \\
            \texttt{deepmatcher}~\cite{deepmatcher2018}                    & \ding{56}   & \checkmark & \href{https://raw.githubusercontent.com/anhaidgroup/deepmatcher/refs/heads/master/deepmatcher/runner.py}{\texttt{\scriptsize{deepmatcher.runner}}}                                                                               \\
            \href{https://github.com/larsga/Duke}{Duke}                    & \checkmark  & \checkmark & \href{https://github.com/larsga/Duke/blob/master/duke-core/src/main/java/no/priv/garshol/duke/Duke.java}{\texttt{\scriptsize{duke.Duke}}}                                                                                        \\
                                                                           &             &            & \href{https://github.com/larsga/Duke/tree/master/duke-server/src/main/java/no/priv/garshol/duke/server}{\texttt{\scriptsize{duke.server}}}                                                                                       \\
            JedAI Toolkit~\cite{jedai2018}                                 & \ding{56}   & \checkmark & \href{https://github.com/scify/JedAIToolkit/tree/134869c94299bf40877097cce1075b127b18650e/src/main/java/org/scify/jedai/workflowbuilder}{\texttt{\scriptsize{jedai.workflowbuilder}}}                                            \\
            FAMER~\cite{famer2018}                                         & \checkmark  & \checkmark & \href{https://git.informatik.uni-leipzig.de/dbs/FAMER/-/blob/master/famer-configuration/src/main/java/org/gradoop/famer/configuration/FamerConfiguration.java}{\texttt{\scriptsize{FamerConfiguration}}}                         \\
                                                                           &             &            & \href{https://git.informatik.uni-leipzig.de/dbs/FAMER/-/blob/master/famer-configuration/src/main/java/org/gradoop/famer/configuration/incremental/IncrementalConfiguration.java}{\texttt{\scriptsize{IncrementalConfiguration}}} \\
            Splink~\cite{splink2022}                                       & \ding{56}   & \checkmark & \href{https://github.com/moj-analytical-services/splink/blob/master/splink/internals/database\_api.py}{\texttt{\scriptsize{internals.database\_api}}}                                                                            \\
                                                                           &             &            & \href{https://github.com/moj-analytical-services/splink/blob/master/splink/internals/linker.py}{\texttt{\scriptsize{internals.linker}}}                                                                                          \\
            d-blink~\cite{dblink2021}                                      & \ding{56}   & \checkmark & \href{https://github.com/cleanzr/dblink/blob/master/src/main/scala/com/github/cleanzr/dblink/Run.scala}{\texttt{\scriptsize{dblink.Run}}}                                                                                        \\
        \end{tabularx}
        \caption{
            Incremental and batch capability of the reference ER systems.
        }
    \end{center}\label{tbl-er-system-inc-batch}
\end{table}%

\begin{figure}[htbp]
    \begin{center}
        \includegraphics[width=\textwidth]{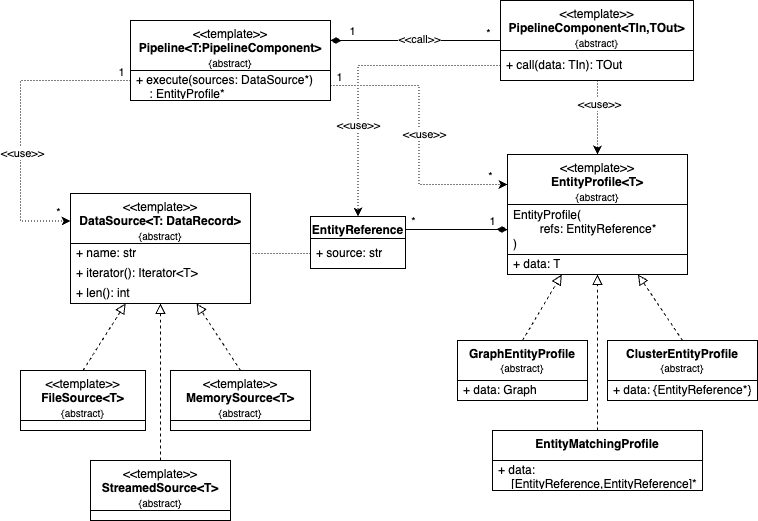}
        \caption{
            Top level ER abstractions which set the framework for ER, as
            described in Subsection~\ref{subsec-information-flow}.
        }\label{fig-lv-top-level-participants}
    \end{center}
\end{figure}

Regardless of specific implementation details, entity resolution fundamentally
follows a pipeline model, transforming input records into structured entity
profiles.
According to the original pipeline design pattern, the most effective approach
to pipeline construction is component-based design.
Figure~\ref{fig-lv-top-level-participants} illustrates key top-level
abstractions common to all reference ER systems, though they may appear in
different forms in practice.
The diagram emphasizes the diversity of input sources and the various ways
output data can be represented.
Additionally, it serves as a reminder that pipeline components primarily process
entity references and entity profiles.
This characteristic structure of the ER pipeline enhances system
comprehensibility, particularly for newcomers.
The key takeaway is that once entity references have been determined within each
data source, they can be processed iteratively to generate entity profiles in
the desired format using a series of pipeline components.

The following subsections provide more detailed descriptions for key pipeline
components from Figures~\ref{fig-lv-batch-er} and~\ref{fig-lv-incremental-er}.

\subsection{Entity Reference Extraction}\label{subsec-lv-eref-extraction}

Information in an ER system undergoes a series of transformations, the first of
which is arguably the most impactful, as it standardizes input across diverse
data sources.
This process, sometimes referred to as data cleaning~\cite{e2e2020,blocking2020,dblink2021},
is known here as entity reference extraction and is responsible for converting
raw information records into structured entity references.

\begin{figure}[htbp]
    \begin{center}
        \includegraphics[width=\textwidth]{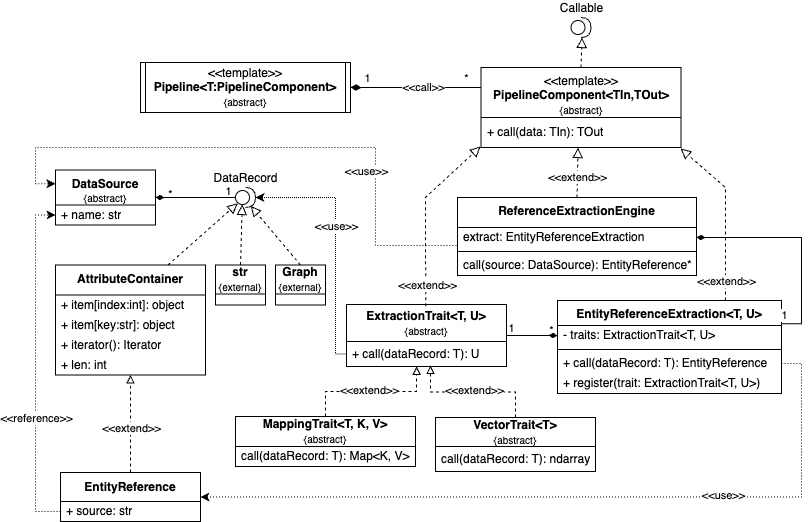}
        \caption{
            The entity reference extraction subsystem, described in
            Subsection~\ref{subsec-lv-eref-extraction}, consists of multiple
            logical components that transform any type of input into a
            structured collection of entity references.
        }\label{fig-lv-er-extraction}
    \end{center}
\end{figure}

Broadly, entity reference extraction follows two dominant paradigms.
Some systems favor extracting vectors from the input information, producing
numerical representations suitable for training or integrating with neural
networks.
More traditionally, the extracted information is structured into objects with
attributes.
Note that such objects include node vertices, mapping types or columnar storage
data types~(e.g dataframes).
Table~\ref{tbl-er-system-extraction} shows this entity reference extraction
categorization for our selection reference systems.

\begin{table}[htp]
    \begin{center}
        \begin{tabularx}{\textwidth}{lcl}
            System                                                         & Representation & Source Code                                                                                                                                                                                         \\
            \toprule
            \texttt{ditto}~\cite{ditto2020} + Magellan~\cite{magellan2018} & Attributes     & \href{https://github.com/anhaidgroup/py\_entitymatching/tree/master/py\_entitymatching/io}{\texttt{py\_entitymatching.io}}                                                                          \\
                                                                           &                & \href{https://github.com/megagonlabs/ditto/blob/master/ditto\_light/dataset.py}{\texttt{ditto\_light.dataset}}                                                                                      \\
                                                                           &                & \href{https://github.com/megagonlabs/ditto/blob/master/ditto\_light/augment.py}{\texttt{ditto\_light.augmenter}}                                                                                    \\
            \texttt{deepmatcher}~\cite{deepmatcher2018}                    & Attributes,    & \href{https://github.com/anhaidgroup/py\_entitymatching/tree/master/py\_entitymatching/io}{\texttt{py\_entitymatching.io}}                                                                          \\
                                                                           & Vectors,       & \href{https://github.com/anhaidgroup/deepmatcher/blob/master/deepmatcher/models/attr\_summarizers.py}{\texttt{attr\_summarizers}}                                                                   \\
                                                                           & Hybrid         &                                                                                                                                                                                                     \\
            \href{https://github.com/larsga/Duke}{Duke}                    & Attributes     & \href{https://github.com/larsga/Duke/tree/b561c409ae455f13f38556fabb6656825ffb7e2e/duke-core/src/main/java/no/priv/garshol/duke/datasources}{\texttt{datasources}}                                  \\
                                                                           & N-Triples      & \href{https://raw.githubusercontent.com/larsga/Duke/b561c409ae455f13f38556fabb6656825ffb7e2e/duke-lucene/src/main/java/no/priv/garshol/duke/databases/LuceneDatabase.java}{\texttt{LuceneDatabase}} \\
            JedAI Toolkit~\cite{jedai2018}                                 & Attributes     & \href{https://github.com/scify/JedAIToolkit/tree/master/src/main/java/org/scify/jedai/datareader}{\texttt{jedai.datareader}}                                                                        \\
                                                                           &                & \href{https://github.com/scify/JedAIToolkit/blob/master/src/main/java/org/scify/jedai/datamodel/}{\texttt{jedai.datamodel}}                                                                         \\
            FAMER~\cite{famer2018}                                         & Graph Vertices & \href{https://git.informatik.uni-leipzig.de/dbs/FAMER/-/tree/master/famer-preprocessing/src/main/java/org/gradoop/famer/preprocessing/io}{\texttt{famer.preprocessing.io}}                          \\
            Splink~\cite{splink2022}                                       & Attributes     & \href{https://github.com/moj-analytical-services/splink/blob/master/splink/internals/database\_api.py}{\texttt{splink.internals.database\_api}}                                                     \\
                                                                           &                & \href{https://github.com/moj-analytical-services/splink/blob/master/splink/internals/splink\_dataframe.py}{\texttt{splink.internals.splink\_dataframe}}                                             \\
            d-blink~\cite{dblink2021}                                      & Attributes     & \href{https://github.com/cleanzr/dblink/blob/dc3dd0daf55f8a303863423817a0f0042b3c275a/src/main/scala/com/github/cleanzr/dblink/Project.scala\#L175}{\texttt{dblink.Project}}                        \\
        \end{tabularx}
        \caption{
            Entity reference extraction~(described in Subsection~\ref{subsec-lv-eref-extraction})\@
            implementations in the reference systems.
        }
    \end{center}\label{tbl-er-system-extraction}
\end{table}%

Outside of our selection of reference ER systems, a particularly interesting
case of attribute-based extraction arises in Named Entity Recognition and
Disambiguation (NERD) systems.
Frameworks such as spaCy~\cite{spacy2023} and NLTK~\cite{nltk2004} extract named
entities from unstructured text while also providing associated attributes.
This structured output makes NERD-derived data structuring similar to that used
in classical record linkage tasks.

In the case of entity alignment the extracted data is structured as a knowledge
graph~\cite{eakit2021}.
Unlike graph-based record linkage implemented by FAMER\cite{famer2018}, which
primarily stores similarity relationships between entity references, entity
alignment operates over knowledge graphs, identifying correspondences across
different graphs.
While both methods generate graph representations, their treatment in the ER
pipeline differs: entity alignment constructs separate knowledge graphs per
each data source, whereas graph-based record linkage represents each entity
reference as a vertex, integrating all references into a single logical graph.
This divergence from the typical record linkage logic must first be addressed by
entity alignment system architectures in the entity reference extraction
component.

Much of the terminology introduced in earlier sections of this reference
architecture is rooted in entity reference extraction.
Figure~\ref{fig-lv-er-extraction} outlines general abstractions derived from
all reference ER systems.
While implementation details vary, every reference system includes an extraction
engine which orchestrates data retrieval and transformation into an internal
representation.
The specifics of this format are left open and differ widely in practice.

Entity reference extraction may rely on abstractions we refer to as extraction
traits.
These are flexible, extensible algorithms that define the format, function, and
meaning of extracted data.
Traits can be simple, extracting fundamental data types like floating-point
numbers or keywords~(e.g~\texttt{dblink.Project} CSV parsing), or compound,
combining multiple traits~(e.g~\texttt{ditto\_light.augment} module which
allows performing several chained operations on the data).
The main purpose of extraction traits is to adapt data representations to
internal ones.
Making extraction traits part of an ER system's architecture allows tailoring
its data ingestion to various use cases while maintaining a consistent pipeline
for downstream ER tasks.

Despite methodological differences, all approaches ultimately produce an
instance of an internal data structure specific to each ER system.
The instance produced by an assortment of extraction traits is referred to as an
entity reference.
The data structure useful in ER contains one or more entity references.
An admitted limitation of the reference architecture is the inability to specify
which data structure is most suitable.
This limitation stems from the multiple available options, the choice between
which is driven by the ER model choice.
For example, in record linkage scenarios we might use an algebraic set whereas
entity alingment requires using knowledge graphs.

Multiple design patterns are often evidentiated in extraction trait design, with
the result being most frequently recognized as adapters, builders or composite
objects~\cite{gofdesign1994}.
Designing families of extraction traits might also be of interest because of
technology constraints (e.g multiple data source types).
Such designs are often recognized as abstract factories~\cite{gofdesign1994}.

Once entity references have been extracted, they serve as the foundation for
subsequent ER phases, ensuring that all downstream transformations operate on a
consistent representation of input data.

\subsection{Comparison Space Generator}\label{subsec-lv-comparison-space}

The computational complexity of entity resolution (ER) scales quadratically with
the total number of entity references in the system, as every entity reference
must be compared to every other.
In incremental ER, this issue appears mitigated due to the asymmetry between the
input record size and the internal entity reference database.
However, Simonini et al.~\cite{schemaagnosticproger2019} emphasize that
incremental ER must be fast and cannot afford exhaustive matching across an
entire large database.
Instead, it relies on blocking and neighborhood indexing to restrict comparisons
to a small, relevant subset of records.
The methods introduced in their work (PSN, PBS, PPS) prioritize comparisons
efficiently, ensuring that entity clusters form incrementally without requiring
exhaustive pairwise matching.

Another observation on existing systems is that a reliance on both matching and
clustering is not necessary for performing ER.\@
Alternatives exist which bypass either matching\cite{dblink2021} or
clustering\cite{deepmatcher2018}.
One might assume that omitting the traditional matching step removes the need to
constrain the number of comparisons.
However, the fundamental challenge of a quadratic comparison space remains in
all cases, as pointed out by previous literature reviews~\cite{blocking2020}.

\begin{figure}[htbp]
    \begin{center}
        \includegraphics[width=\textwidth]{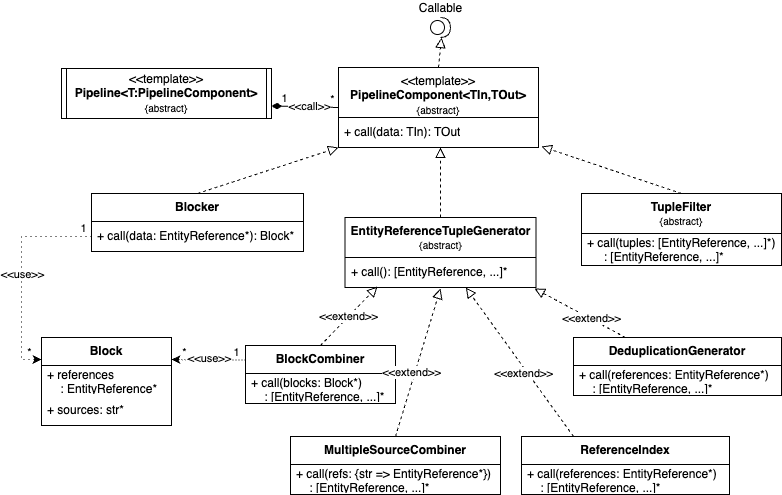}
        \caption{
            Comparison space generation yields collection of entity references
            through various methods such as blocking and indexing, described in
            Subsection~\ref{subsec-lv-comparison-space}.
        }\label{fig-lv-comparison-space}
    \end{center}
\end{figure}

Figure~\ref{fig-lv-comparison-space} should be regarded as a straw man which
highlights key mechanisms for controlling the comparison space within the ER
runtime.
Methods such as blocking, indexing, and filtering serve to reduce the number of
comparisons that must be performed further in the ER pipeline.

\begin{table}[htp]
    \begin{center}
        \begin{tabularx}{\textwidth}{lcccl}
            System                                                         & B          & I          & F          & Source Code                                                                                                                                                                                                           \\
            \toprule
            \texttt{ditto}~\cite{ditto2020} + Magellan~\cite{magellan2018} & \checkmark & \ding{56}  & \ding{56}  & \href{https://github.com/anhaidgroup/py\_entitymatching/tree/master/py\_entitymatching/blocker}{\texttt{py\_entitymatching.blocker}}                                                                                  \\
            \texttt{deepmatcher}~\cite{deepmatcher2018}                    & \checkmark & \ding{56}  & \ding{56}  & \href{https://github.com/anhaidgroup/py\_entitymatching/tree/master/py\_entitymatching/blocker}{\texttt{py\_entitymatching.blocker}}                                                                                  \\
            \href{https://github.com/larsga/Duke}{Duke}                    & \checkmark & \checkmark & \ding{56}  & \href{https://raw.githubusercontent.com/larsga/Duke/b561c409ae455f13f38556fabb6656825ffb7e2e/duke-core/src/main/java/no/priv/garshol/duke/databases/AbstractBlockingDatabase.java}{\texttt{AbstractBlockingDatabase}} \\
                                                                           &            &            &            & \href{https://raw.githubusercontent.com/larsga/Duke/b561c409ae455f13f38556fabb6656825ffb7e2e/duke-lucene/src/main/java/no/priv/garshol/duke/databases/LuceneDatabase.java}{\texttt{LuceneDatabase}}                   \\
            JedAI Toolkit~\cite{jedai2018}                                 & \checkmark & \checkmark & \checkmark & \href{https://github.com/scify/JedAIToolkit/tree/master/src/main/java/org/scify/jedai/blockbuilding}{\texttt{jedai.blockbuilding}}                                                                                    \\
                                                                           &            &            &            & \href{https://github.com/scify/JedAIToolkit/blob/master/src/main/java/org/scify/jedai/blockprocessing/}{\texttt{jedai.blockprocessing}}                                                                               \\
            FAMER~\cite{famer2018}                                         & \checkmark & \ding{56}  & \checkmark & \href{https://git.informatik.uni-leipzig.de/dbs/FAMER/-/blob/master/famer-linking/src/main/java/org/gradoop/famer/linking/blocking/}{\texttt{famer.linking.blocking}}                                                 \\
                                                                           &            &            &            & \href{https://git.informatik.uni-leipzig.de/dbs/FAMER/-/blob/master/famer-linking/src/main/java/org/gradoop/famer/linking/selection/}{\texttt{linking.selection}}                                                     \\
            Splink~\cite{splink2022}                                       & \checkmark & \ding{56}  & \ding{56}  & \href{https://github.com/moj-analytical-services/splink/blob/master/splink/internals/blocking.py}{\texttt{splink.internals.blocking}}                                                                                 \\
            d-blink~\cite{dblink2021}                                      & \ding{56}  & \checkmark & \ding{56}  & \href{https://github.com/cleanzr/dblink/tree/master/src/main/scala/com/github/cleanzr/dblink/partitioning}{\texttt{dblink.partitioning}}                                                                              \\
        \end{tabularx}
        \caption{
            Comparison space generation techniques supported by the reference
            systems.~\textbf{B} stands for \emph{blocking}, \textbf{I} stands
            for \emph{indexing} and \textbf{F} stands for comparison
            \emph{filtering}.
        }
    \end{center}\label{tbl-er-system-comparison-generation}
\end{table}%

Blocking, indexing and filtering are collectively referred to as comparison
space generation methods from here on.
Table~\ref{tbl-er-system-comparison-generation} lists how each of the reference
systems handles comparison space generation.

As with entity reference extraction, there is a limit to the amount of details
that can be specified in the reference architecture.
For example, while tempting to imagine modelling comparison space generation
methods using a coherent unitary contract, this is not necessarily possible.
The JedAI system provides a good example in this regard.
Its~\texttt{org.scify.jedai.blockprocessing} package defines several
filtering and indexing implementations which expect as input the output of a
blocking operation.
Therefore, detailing the design of individual comparison space generators does
not seem wise.

On the other hand, it is common in practice for pairs of entity references to be
output by the comparison generation subsystem.
This can be seen in practice across all reference systems, including FAMER
(which outputs pairs of graph vertices).
However, in Figure~\ref{fig-lv-comparison-space} we envision the posibility that
multiple entity references participate in subsequent comparisons.
The reference architecture is suitable for models which generalize ER matching
to this effect~\cite{multier2013} because of this.
This provides an extension point for using such models.

Finally, implementing system architectures may wish to allow storing the
comparison space in a dedicated reference store to reduce the running time of
subsequent resolutions over the same data or to support incremental ER.\@

\subsection{Reference Store}\label{subsec-lv-reference-store}

The Reference Store appears in both incremental and batch ER, even though it is
strictly necessary only for the incremental variant.
Existing systems often implement it as an in-memory structure.
Common representations include a sequence of pairs of matching entity references
(e.g Ditto~\cite{ditto2020}), a similarity graph~(e.g. JedAI~\cite{jedai2018})
or a logical graph~(e.g FAMER~\cite{famer2018}).
Certain closed-source implementations reportedly persist long-lived identifiers
linked to entity references~\cite{fusion2020}, though their exact design remains
unknown.

This component plays the central role in entity alignment where it is synonymous
with either the source or the target knowledge graph~\cite{eakit2021,openea2020}.
For record linkage implementations, Duke's~\texttt{Database} components provide
a flexible example of a reference store.
Splink's use of Apache Spark tables to store and quickly retrieve relevant data
exemplified in its \texttt{blocking} module is another example which showcases
how the reference store logical abstraction might be designed in practice.

When it is designed as a component, it is reminiscent of the \emph{Memento}
design pattern~\cite{gofdesign1994} because it is primarily used to store and
access versions of the same data.
Common concerns to consider when designing the reference store include data
availability and integrity.

\subsection{Matching Engine}

The vast majority of the reference ER systems we examined include a matching
subsystem as part of their batch ER pipelines.
The matching engine represents a black-box component, and there are two key
reasons for this.

Existing ER systems employ a range of matching approaches, including
probabilistic matching\cite{splink2022}, deterministic rule-based
matching\cite{jedai2018}, neural network-based
matching\cite{deepmatcher2018,ditto2020}, and graph-based
methods\cite{famer2018}.
Each of these models expects different input formats and produces distinct
outputs.
Furthermore, matching is a highly active area of research, leading to diverse
methodologies that vary significantly across implementations.
As such, defining the matching subsystem is left open.

Additionally, the matching process can be symbolically reduced to a functional
representation, barring the need to inspect the detailed logical components
involved.
The black box approach to matching was popularized by the Stanford Entity
Resolution Framework (SERF)~\cite{serf2009}.
One critique of the SERF match function is that it implicitly assumes a fixed
output structure, whereas in practice, matching strategies differ widely.

While matching and clustering often work together in ER, they are not strictly
dependent on each other.
At least one must be present to resolve entities.
Matching reflects the fundamental way in which an observer—human or
algorithmic—identifies similarity between data points and prior knowledge.
The presence of a matching subsystem in the ER pipeline has pragmatic benefits.
For example, a matching subsystem can produce a similarity graph which captures
similarity relationships between entities, resulting in more nuanced clustering
strategies downstream~\cite{jedai2018,famer2018}.
Additionally, due to a lower amount of information presented to the user,
interfaces highlighting individual pairs of potentially matching entity
refernces might be easier to build and easier to make user friendly.

System architectures should design the matching engine in the most suitable
way to their problem domain and chosen ER model.
Certain architectures may benefit from leveraging multiple matching engines.
A legitimate concern may regard the ability to scale the matching engine
independently of the other components in the ER pipeline because of the
higher computational expense inherent in this component.
These considerations encourage using the \emph{Strategy} design
pattern~\cite{gofdesign1994} in object-oriented designs or functional
programming for encapsulating the algorithms which sit at the core of the
matching engine.
The concerns around scalability and interoperability of the matching engine with
its surrounding pipeline components strongly suggest using a component-oriented
design.

Finally, the possibility to implement systems that do not leverage similarity,
renders the matching engine as an optional component in the ER pipeline.

\subsection{Clustering Engine}

Clustering plays a crucial role in entity resolution (ER) by grouping linked
entity references that correspond to the same real-world object.
It follows the principle that things which co-occur or exhibit strong relational
ties likely refer to the same underlying entity.
This mechanism is complementary to matching which uses the general concept of
similarity to provide the evidence of solid relational ties.
Both mechanisms contribute to ER in an essential way, though only at least one
of them is strictly necessary.

Widely regarded as an essential step, not all reference systems implement it
transparently.
Some ER systems, such as Duke~\cite{duke2014} obfuscate clustering by making it
part of intermediate storage operations which use its system of
\texttt{MatcherListener}s.
System developers are informed about clustering possibilities only once the
\texttt{LinkDatabaseMatchListener} and \texttt{LinkDatabase} concepts are
perused.
This obfuscates the pipeline structure of the ER process.
On the other hand, some entity matching systems such as Ditto~\cite{ditto2020}
or DeepMatcher~\cite{deepmatcher2018} focus solely on identifying matching
entity references, partially or completely omitting transitive resolution.
In some cases, this approach can be beneficial, as it offloads the
\textit{bad-triplet} problem~\cite{badtriplet2008} to downstream systems.

At the other end of the spectrum, entity alignment and record linkage systems
designed around graph models inherently support clustering.
Some ER solutions go further by eliminating the matching phase entirely.
For example, clustering subsumes matching by employing functional similarity
measures and a carefully chosen merge function~(e.g SERF~\cite{serf2009}),
generic distance metrics~\cite{pretrainedembeddingclusters2023}, or
probabilistic models which form clusters directly from the comparison space
using similarity functions to model colocation~(e.g D-blink~\cite{dblink2021}).

When applied, clustering typically operates on data supplied by the reference
store.
In cases where matching is omitted, the entity reference store effectively
defines the relevant comparison space.
Systems such as the JedAI~\cite{jedai2018}
(\href{https://github.com/scify/JedAIToolkit/tree/134869c94299bf40877097cce1075b127b18650e/src/main/java/org/scify/jedai/entityclustering}{\texttt{entityclustering}}
package) and FAMER~\cite{famer2018}~(\href{https://git.informatik.uni-leipzig.de/dbs/FAMER/-/tree/master/famer-clustering/src/main/java/org/gradoop/famer/clustering?ref_type=heads}{\texttt{famer-clustering}}
package) already support various clustering
algorithms~\cite{bipartitegraphclusters2023,compclusteringeval2017}, offering
flexible options for different ER workflows.

Given its varying adoption across reference systems and the limited research on
cluster coherence~\cite{fusion2020}, system architectures may design a
clustering engine if their specific problem domain and ER model of choice
require one.
In object-oriented designs, the \emph{Strategy} design
pattern~\cite{gofdesign1994} may be used to encapsulate the various algorithms
sitting at the core of the clustering engine.
Similarly to the matching engine, component oriented design improves
interoperability and, indirectly, scalability.

New system architectures should carefully consider clustering engines for
incremental ER.\@
Only some clustering algorithms~(such as agglomerative clustering)~are generally
incremental, i.e can be easily carried out using the batch linkage
method~\cite{incrrl2014}.

Finally, the evaluation of clustering performance is not as common among the
reference ER systems as the fundamental importance of the clustering subsystem
to ER would imply.
D-blink~\cite{dblink2021} is an example system which implements this feature by
implementing the Adjusted Rand Index~\cite{adjrand2001} for evaluating ER
quality.
ER system architectures which incorporate clustering engines should also provide
the means for evaluating clustering performance along with matching performance,
especially since libraries which implement clustering metrics for ER are readily
available~\cite{pyresolvemetrics2024}.
Future efforts in this area would improve ER system transparency and
reliability.

\subsection{Presentation Layer}\label{subsec-lv-presentation}

The logical component found across all ER systems is the entity profile
presentation layer which has the responsibility of making \emph{entity profiles}
available to ER consumers.
While entity profile formatting or display vary according to the needs of each
implementation, all ER systems define a logical component which assembles entity
profiles.
The data representation of entity profiles varies along with the chosen
approach to ER.
The act of designing ER system architectures necessarily involves choosing an ER
model which best suits the problem domain.
The entity profile assembler logic will necessarily be dictated by the selected
ER model.

\begin{figure}[htbp]
    \begin{center}
        \includegraphics[width=\textwidth]{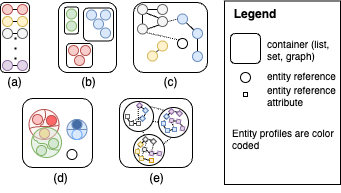}
        \caption{
            Examples of entity profile representations.
            In entity matching models, a pair of entity references often
            constitutes the entity profile (a).
            Traditional clustering-based approaches resolve transitivity but
            they typically assign each entity reference to a single cluster (b).
            A logical or similarity graph generalizes this further, allowing
            entity profiles to emerge as subgraphs where an entity reference may
            belong to multiple profiles (c).
            More advanced models, such as SERF’s merge dominance framework (d)
            and entity alignment models (e), construct entity profiles by
            merging attributes from entity references.
            The SERF model treats entity profiles as entity references
            themselves, enabling incremental refinement, while entity alignment
            follows a knowledge graph ontology, reducing the importance of
            original entity references.
            Subsection~\ref{subsec-lv-presentation} describes the role of these
            representations in the reference architecture.
        }\label{fig-lv-entity-profiles}
    \end{center}
\end{figure}

This poses a problem for the interoperability of ER systems because of the
myriad ways to represent entity profiles used in practice.
Ditto represents entity profiles as pairs of matching entity
references~(Ditto~\cite{ditto2020}).
Entity profiles take the form of algebraic set partitions when using Duke's
equivalence class linker~\cite{duke2014}, JedAI's \texttt{ExactClustering}
component~\cite{jedai2018} or D-blink's default output~\cite{dblink2021}.
FAMER represents entity profiles as subgraphs~\cite{famer2018}.
Similarly, entity alignment~\cite{openea2020} systems also represent entity
profiles as graphs.
The main distinction is that in this case entity profiles are created guided
by the knowledge graph's ontology rather than by relationships between entity
references.

SERF~\cite{serf2009} introduces a black-box merge function that constructs more
\emph{dominant} entity references incrementally using the matching engine's
output.
The R-Swoosh algorithm which defines this approach is exemplified in practice by
the Oyster~\cite{oyster2013} ER system.
The final entity references constructed in this way represent the entity
profiles, their data structure being determined flexibly by the black box merge
function.
Merge dominance logically enables the use of the ER pipeline's output as entity
references in subsequent runs.
Although treating entity profiles as entity references provides flexibility, it
also introduces challenges.
The traceability and explainability of ER may be reduced, as subsequent links
derived from newly generated entity profiles could obscure provenance, affecting
auditability and trust.
Additionally, if entity clusters lack coherence, erroneous merges may propagate
downstream, exacerbating the bad-triplet problem.

The above aspects highlight the need for careful selection of entity profile
representations.
Structured attributes, clusters, and merged profiles each present trade-offs in
interpretability and usability.
Preserving traceability when merging references is critical for explainable ER,
particularly in regulated domains favouring representations of ER profiles which
are more easily deconstructible to source enity references, such as clusters or
graphs.
Ensuring cluster coherence before treating entity profiles as entity references
helps prevent iterative resolution errors.

In terms of output quantity, there is a key distinction between incremental ER
and batch ER.\@
Incremental ER returns only entity profiles based on records up until the latest
input~\cite{incrrl2014}, perhaps even focused on the latest record.
In contrast, batch ER and entity alignment return all resolved entity profiles
at once.
This distinction affects how results are consumed: incremental ER may provide
focused, query-driven resolution, whereas batch ER enables global reconciliation
of all records.

Viewing incremental ER as a query-driven system can hint at possible ways to
solve the interoperability problem mentioned earlier in this section.
A component oriented design coupled with a web service architecture might
provide the standardization required to make ER systems more interoperable.

A well set up reference store mitigates many of the concerns related to entity
profile usage and representation.
Entity profile design and its integration into the ER pipeline—particularly
in incremental ER—should be guided by the specific problem domain.
Key motivations include provenance tracking, structured representations,
security and privacy considerations, and scalable entity aggregation.

\section{ER Runtime Process}\label{sec-pv}

Up until this point, the paper focused on generic use cases and abstract logical
components.
Procedurally, ER may be interpreted as a series of subprocesses and activities
which, in abstract, remain consistent across existing approaches.

\begin{sidewaysfigure}[htbp]
    \begin{center}
        \includegraphics[width=\textwidth]{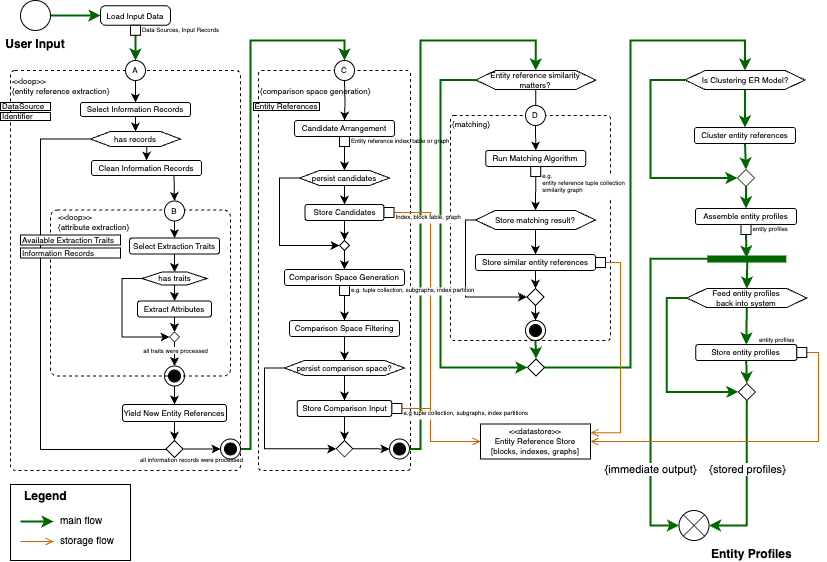}
        \caption{
            Entity Resolution Process described in Section~\ref{sec-pv}.
            The process begins with loading input data, which may originate from
            multiple data sources (in batch ER) or be ingested incrementally.
            Next, entity reference extraction (A) converts information records
            into structured entity references, using traits (B) to extract
            relevant attributes.
            Once entity references are generated, the comparison space (C) is
            constructed to reduce the computational complexity of ER.\@
            If resolution is achieved through similarity-based analysis, the
            system performs matching (D).
            Otherwise, it proceeds directly to clustering and assembling entity
            profiles.
            Optionally, resolved entity profiles can be fed back into the system
            by storing them in the entity reference store.
            The entity reference store may also retain other intermediate
            artifacts throughout the process.
            The process concludes by returning the final entity profiles.
        }\label{fig-pv-main}
    \end{center}
\end{sidewaysfigure}

Figure~\ref{fig-pv-main} presents a generalized blueprint of the ER pipeline,
capturing its fundamental stages.
It also emphasizes key subprocesses through capital-letter annotations.

This blueprint should not be interpreted as a fixed sequence of algorithmic steps.
As noted in Section~\ref{sec-lv}, ER runtimes exhibit
significant variability, ranging from implementations closely aligned with this
conceptual model to architectures based on SERF~\cite{serf2009} or end-to-end
neural network approaches~\cite{deepmatcher2018}.
While the blueprint serves as a reference for identifying common structural
components across ER systems, the ultimate design and composition of an ER
pipeline remain at the discretion of system architects.

\subsection{Entity Reference Extraction}\label{subsec-pv-eref-extraction}

Entity resolution systems differ in their representation of entity references,
which may take the form of tuples~\cite{deepmatcher2018},
mappings~\cite{duke2014,splink2022}, sets of named attributes~\cite{jedai2018},
vectors~\cite{ditto2020}, or vertices within a logical graph~\cite{famer2018}.
Regardless of the chosen format, structuring data is a fundamental step in the
ER process.
It ensures that the input is represented in a consistent and processable manner.

The extraction process~\textbf{(A)} begins with the selection of relevant
information from the input data.
This data may be introduced incrementally or in batch.
Using source identifiers to indicate provenance is a common approach~\cite{deepmatcher2018,jedai2018}.
Unlike existing systems that rely on structured input formats, such as treating
each row of a CSV file as a distinct information record, this approach
generalizes the extraction process to accommodate unstructured data sources.
Named Entity Recognition and Disambiguation (NERD) systems~\cite{nerd2012}
exemplify how information records may be derived without explicit schema
constraints.
Once identified, the extracted records usually undergo a cleaning phase, where
inconsistencies are resolved, and redundant or irrelevant data is removed.
This ensures a standardized quality across extracted information.

Following data selection and cleaning, it is customary to extract
attributes~\textbf{(B)}.
This subprocess may take one or more clean information records as input.
While systems typically process information records one at a time (e.g lines in
a CSV file), there are cases when multiple information records are relevant
together~(e.g scraping a web page for cohesive information).
Because each inspected system has its own attribute extraction mechanisms, we
introduce a common abstraction called \textit{extraction traits} to generalize
attribute extraction.
Extraction traits are predefined algorithms identify meaningful attributes in
the cleaned information records.
In that, extraction traits are a form of \emph{Adapters}~\cite{gofdesign1994}.
Extraction traits may also be \emph{Composite}~\cite{gofdesign1994} objects,
thus enriching the attribute extraction capabilities of the ER system.
Each extraction trait may yield a collection of attributes, with system-specific
logic.
Extraction traits should be allowed to determine whether records are skipped,
merged, or split, too.
These transformations are treated as black-box processes, as the reference
architecture can not impose constraints on the internal logic of extraction
algorithms.
The only recommendation is that the process operate over a positive number
of input records and return attributes that partially define an entity
reference.
The attribute extraction subprocess~\textbf{(B)} ends when all predefined
extraction traits were employed to process the information records.

The entity reference extraction process~\textbf{(A)} continues with the
construction of entity references.
This activity naturally follows the attribute extraction process.
Depending on system requirements, entity references may be formed by
concatenating extracted attributes~\cite{jedai2018}.
More complex setups integrate entity reference construction in more complex
designs, such as neural network architectures~\cite{deepmatcher2018}.
While this framework is well suited for entity references structured as mappings
or tuples, its usage in vector-based representations requires adaptation.
In such cases, entity references may either be treated as mappings with one or
more vector attributes or be defined as a fixed set of attributes corresponding
to the vector’s dimensions.

Once entity references are assembled, the process continues iteratively until no
further input data remains.
This allows enough flexibility to design batch, incremental, and online ER
systems.

When the input data shares the same structure as the internal representation of
entity references, the extraction process may trivially allow data to pass
through without modification.

\subsection{Comparison Space Generation}\label{subsec-pv-csg}

In its basic form ER has a time complexity of $\mathcal{O}(n^2)$~\cite{e2e2020},
where $n$ represents the total number of entity references.
This complexity arises from the need to compare all entity references among
themselves, making the approach infeasible for large-scale data.
To mitigate this, blocking, indexing, and filtering techniques~\cite{blocking2020}
reduce the number of comparisons while preserving resolution quality.
These techniques benefit both batch and incremental ER by improving processing
efficiency.

Since the number and type of comparison-reducing methods vary across
implementations, a flexible abstraction, termed the \emph{comparison space}, is
introduced.
This concept organizes entity references in structured ways that facilitate
minimizing the number of required similarity comparisons while retaining
essential relationships.

The process of generating the comparison space, denoted as \textbf{(C)} in
Figure~\ref{fig-pv-main}, consists of three main phases: candidate arrangement,
comparison input generation, and filtering.
These phases largely echo the JedAI Toolkit's process of blocking, block
cleaning and comparison cleaning.
Candidate arrangement reorganizes entity references into efficient data
structures without loss of information.
Techniques such as blocking, indexing, and graph edge modification exemplify
this step, resulting in structures like block tables, balanced trees, or graphs.
In some cases, these structures may be persistently stored for later retrieval.

Once candidate arrangement is complete, comparison input generation produces
reference pairs or groups suitable for the subsequent matching phase.
Since matching algorithms operate on predefined input parameters, this step
determines the number of comparisons that will be executed.

Filtering further refines the comparison space by eliminating low-relevance
entity pairs before matching.
While this step enhances efficiency, it may introduce information loss if
certain entity references are entirely excluded~\cite{e2e2020}.

The final comparison space, potentially stored for optimization, allows
flexibility in tuning matching and clustering strategies.
This structured approach ensures that ER systems scale efficiently without
compromising accuracy.

As a final remark, the actions for storing the ``comparison space'' might
resemble the \emph{Memento} design pattern~\cite{gofdesign1994}.

\subsection{Matching and Clustering}

Entity resolution fundamentally relies on two core components: matching and
clustering.

Matching, represented abstractly in Figure~\ref{fig-pv-main} as
subprocess~\textbf{(D)}, identifies references to the same entity by evaluating
their similarity.
This process remains a black box within the reference architecture due to the
diversity of available techniques.
Traditional approaches range from probabilistic and deterministic methods to
state-of-the-art deep learning models, including transformers~\cite{ditto2020}
and graph neural networks~\cite{eakit2021}.
As the most extensively studied aspect of entity resolution, matching remains at
the forefront of research and system optimization.

Clustering, in contrast, takes advantage of entity references being organized
within a metric space or a graph, leveraging spatial proximity or graph theory
to infer co-referential relationships.
Though historically receiving less attention than matching in ER research,
it has become more studied lately~\cite{dblink2021,pretrainedembeddingclusters2023}.
As with matching, clustering is a black box within the reference architecture
due to the number of options available.

While robust ER solutions integrate both matching and clustering, only one of
these components is strictly necessary.
The Swoosh family of algorithms~\cite{serf2009} allows assigning various degrees
of importance to matching and clustering by carefully choosing the match and
merge functions.
Systems that employ both matching and clustering may benefit from storing
matching results.
Similarity graphs~\cite{jedai2018} and logical graphs~\cite{famer2018} may
provide good starting points for representing entity reference relationships,
facilitating further processing and analysis.

From both a logical and deployment perspective, the entity reference store
serves as a natural repository for storing matching output, ensuring efficient
retrieval and refinement into clusters.

\subsection{Entity Profile Assembly}

The last step of the ER process consists in building the entity profile.
Entity profiles are built from entity reference clusters or directly from the
output of the matching subprocess.

Because of the large number of ways of aggregating entity references into a
coherent entity profile, entity profile assembly system designs might end up
resembling \emph{Builder}s~\cite{gofdesign1994}.

Certain ER models~(like SERF~\cite{serf2009}) allow entity profiles to be viewed
as entity references.
This allows feeding entity profiles back into the system.
The reference store~(Subsection~\ref{subsec-lv-reference-store}) most naturally
addresses this need.

\section{Implementation and Deployment Guidelines}

\subsection{Development Considerations}

Even though every implementation is unique, the inspected reference systems
exhibit idiomatic programming techniques and favor abstraction over hardcoded
logic.
This approach improves system maintainability, and indirectly extensibility,
allowing support for diverse data formats and algorithms which drive the ER
process.
Separating the user interface from the core ER system is beneficial, especially
when the UI evolves independently.
FAMER~\cite{famer2018} and JedAI~\cite{jedai2018} exemplify this approach in
practice.
Similarly, modularizing ER phases enables independent evolution and scalability
of components.

Systems that implement a clear and structured pipeline (e.g., FAMER, JedAI,
Splink, dblink) are more maintainable than those lacking an idiomatic runtime
configuration (e.g., Ditto, DeepMatcher) or those which apply patterns in
unconventional ways.
For instance, Duke employs the \emph{Observer} pattern~\cite{gofdesign1994},
typically used for event-driven programming, to manage workflow execution.
We advocate extending the \emph{Law of Least Astonishment}~\cite{james1987tao}
to source code.
In other words, code should be structured in a way that minimizes surprises for
reviewers.
While this guideline was formulated in a less formal context than literate
programming~\cite{knuth1984literate}, it remains a surprisingly effective
principle for improving code maintainability and readability, thereby increasing
extensibility as well.

Interoperability is an important factor in ER system design, ensuring that both
input data and output results are structured for seamless integration into
larger workflows.
Input records exist in multiple formats, ranging from highly structured
databases to unstructured text.
In multimodal ER, information must be extracted from sources containing
non-textual data, requiring adaptable extraction pipelines.
This necessity highlights the close intermingling of extensibile design
principles and interoperable systems which is so vital for successful end to end
ER systems.

In incremental ER, an API-based approach should be favored, with a clearly
defined query contract for real-time resolution.
A clear separation between input contracts and the internal representation of
entity references is highly encouraged to prevent internal design changes from
affecting the interoperability of the system.
Similarly, entity profiles should be as standardized and portable as possible,
prioritizing machine-readable formats to facilitate downstream integration.
Usability concerns such as human-readable output should remain secondary and
preferably implemented separately.

The requirement to implement custom scaling and parallelization mechanisms
depends almost entirely on the size of the data being processed.
ER systems should leverage existing distributed computing frameworks~(e.g
Dblink~\cite{dblink2021}, Splink~\cite{splink2022} or FAMER~\cite{famer2018}),
reducing the complexity of handling large-scale data processing.

While the JVM, Python, and R dominate current ER implementations, the problem
itself is not bound to any particular technology.
All ER models can be materialized easily using widespread programming paradigms
such as modularization, object orientation or componentization.
Our selection of reference systems clearly exemplifies this possibility.

\subsection{Deployment Considerations}

The scalability needs of a system determine in large part its deployment.
ER systems scale proportionally to the volume of processed entity references.
This is true both of batch and incremental ER systems.
Lightweight deployments are feasible for small datasets or system development,
but scaling the deployment often presents its set of challenges and trade-offs.

Key factors influencing deployment decisions include:

\begin{itemize}
    \item \textbf{Dataset size}: Smaller datasets enable more compact deployments.
          Some systems~\cite{duke2014,jedai2018} can run on a local machine
          without issues.
    \item \textbf{Portability}: Large-scale ER benefits from distributed
          runtimes like Apache Spark~\cite{dblink2021,splink2022} or
          Flink~\cite{famer2018}, but this reduces the portability of the
          system.
    \item \textbf{Distributed computations in incremental ER}: Unlike batch ER,
          incremental ER requires maintaining intermediate state, making
          parallel computations more complex. At the extreme, the CAP theorem~\cite{captheorem2002}
          becomes relevant for the intermediate state storage.
    \item \textbf{Highly available reference store}: Depending on data
          availability and durability requirements, the reference store
          described in Subsection~\ref{subsec-lv-reference-store} may need to be
          deployed in a high-availability (HA) configuration.
\end{itemize}

While scalability often comes at the expense of portability, containerization
(e.g Docker, Kubernetes) and micro virtual machines (e.g AWS Lambda, Azure
functions) can mitigate these limitations by standardizing deployment
environments.
Micro virtual machines may not be suitable for compute-intensive tasks like
batch matching because of resource constraints.
They remain suitable for lower-cost ER phases such as entity reference
extraction.

Beyond scalability and performance considerations, deployment strategies must
align with privacy, security, and regulatory requirements.
ER systems frequently operate in sensitive domains, such as medical record
linkage, where privacy-preserving algorithms and strict data locality policies
are essential.
These setups are favoured by local, bare metal deployments, not cloud native
ones.
On the other hand if global availability, security and auditability concerns
rank highly, cloud native architectures have distinctive advantages.
These auxiliary concerns greatly influence the choice of storage, communication
and data-sharing mechanisms.

\section{Conclusion}

\subsection{Concluding Remarks on Research Questions}

In the preceding sections, we saw that the main use cases of ER systems broadly
fall into three categories: configuration, runtime and evaluation use cases.
Configuration use cases were referred to as the `design time' interface of the
system and highlighted user roles and potential actions that occur when the ER
system is prepared for operation.
In contrast, the runtime use cases highlight the activities surrounding ER
systems which occur while they resolve entity references into entity profiles.
Finally, evaluation use cases highlight activities which are performed either at
design time or at runtime to assess the quality and performance of ER.\@

Then we proceeded by outlining logical~(Section~\ref{sec-lv}) and procedural
(Section~\ref{sec-pv}) abstractions common to all ER systems, regardless of
their focus—record linkage, deduplication, entity alignment, NERD, or other
variations.
These abstractions apply to both batch and incremental ER.\@
Recommendations were provided, and potential pitfalls addressed to ensure that
the logical and procedural components of ER system designs remain extensible,
interoperable, and scalable.
Key design principles, such as loose coupling and high cohesion, were emphasized
alongside established patterns (like the pipeline model) to enhance code
maintainability and readability.
Distributed computing frameworks, including Apache Spark and Flink, were
advocated as preferable alternatives to custom-built runtimes.
From the outset, component orientation was recommended, particularly given the
black-box nature of match and merge operations.

This brings us to the core question: what key insights can we offer for
designing entity resolution systems?
The answer to this question determines the practical effectiveness of the
reference architecture.

A fundamental requirement for ER systems is end-to-end entity resolution.
While we have argued that either matching or clustering is sufficient for ER,
systems should strive to incorporate multiple methods for determining which
entity references contribute to an entity profile.
Our firm recommendation is that ER pipeline components be designed to create
value beyond their individual contributions.
This is achievable only through careful use of abstractions such as the ER
pipeline and the reference store.
Further, batch and incremental ER both provide value, but truly remarkable ER
systems implement both paradigms.
Incremental ER in particular lends itself extremely well to modern architectural
approaches such as service-oriented architecture or event-driven architecture.

A natural consequence of end-to-end ER design is the need for appropriate
evaluation.
Assessing clustering performance solely through statistical metrics (e.g.,
precision, recall, F-measure) leads to information loss.
Conversely, evaluating pairwise clustering performance when no matching has
occurred introduces extraneous data with no meaningful interpretation.
Regardless of the specific combination of matching, clustering, or other
pipeline components, ER evaluation must be implemented and finely tuned to the
runtime environment.

Another critical consideration is the flexibility of the reference architecture.
Configuration, matching, and clustering remain largely at the discretion of
system designers.
While this stance may be viewed as a limitation of the reference architecture,
avoiding the prescription of rigid implementations for these areas is wholly
intentional.
Configuration choices are often tied to the underlying technology, while
matching and clustering depend on the selected ER model.
Furthermore, future methods for selecting entity references beyond matching and
clustering may emerge, and the reference architecture strongly supports
extensibility in this regard.
Matching flexibility allows for deterministic, probabilistic, and
machine-learning-based ER models.
Clustering approaches may include algebraic, probabilistic, or graph-based
methods.
The pipeline design enables seamless substitution of rule-based engines with
neural network models.

Beyond extensibility, large-scale ER requires careful attention to scalability
and performance.
Comparison space generation and reference storage are fundamental abstractions
which must be present in all ER systems.
A component-oriented approach is strongly recommended, as it facilitates
scalability by providing standardized interfaces between component-specific
logic and the overarching pipeline.

ER systems must efficiently process diverse input data while maintaining strict
output consistency to support both upstream and downstream integration.
Logical components which standardize input and output surround a very flexible
and extensible core.
This ensures uniformity in input reduction via comparison space generation as
well as in entity profile assembly based on ER models.
This is possible only because despite differences in implementation, core entity
profile assembly activities remain consistent across ER systems.

These principles lead to a reference architecture that prioritizes
component-oriented design, emphasizing extensibility in configuration, core
runtime components, and evaluation.
Interoperability is achieved by allowing input flexibility through entity
reference extraction while enforcing output standardization through entity
profile assembly.
Scalability and performance are upheld through essential abstractions, including
comparison space generation and the reference store.

Ultimately, this article presents a structured yet adaptable approach to ER
within well-defined boundaries.
The proposed abstractions, principles, and patterns combine themselves
into a reference architecture which can be a structured foundation for future
research and practical implementations.

\subsection{Feedback Process and Validation}

The reference architecture proposed in this paper was developed through a
rigorous study and synthesis of existing systems.
By abstracting their source code into generic concepts and processes, it ensures
empirical soundness.
Efforts were made to maintain a clear correspondence between these abstractions
and their concrete implementations, as outlined throughout the paper.
As such, the architecture provides a structured yet flexible foundation for
designing ER solutions.
However, given the vast scope of existing knowledge, this initial version
remains an imperfect work, at best.

Rather than developing another ER system (which would only contribute to the
paradox of choice aluded to in the introduction), a more effective approach is
to iteratively refine the architecture based on feedback from practitioners.
This transforms it into a dynamic framework, continuously improved using
empirical evaluation and contributions from the ER community.

To support this evolution, researchers and architects are encouraged to
reach out to the authors to establish a collaborative framework.
At a minimum, this approach would allow integrating insights from real-world
implementations, industry constraints, and emerging trends.
The ultimate goal is to sustain an adaptive design that informs and supports the
development of ER systems over time.

\subsection{Final Words}

ER is an area of research with a vast and diverse literature which may cause
a paradox of choice for newcomers to the field.
The present paper introduces a reference architecture for ER focused on
extensibility, interoperability and scalability.
The reference architecture delineates the bounds where generalization and
abstraction yield diminishing returns, particularly for ER configuration and
evaluation, but also for the ER runtime to a lesser extent.
Yet, to our knowledge, this is the first attempt at writing a reference
architecture for ER and the first attempt to unify so many ER flavours using
the same abstractions.

The paper was written in hopes of triggering an effort to standardize ER design
across all flavours of ER.\@
Both subsequent iterations of this reference architecture~(based on feedback and
collaboration), and the standardization of the broader ER problem domain would
be welcome.

\bibliographystyle{elsarticle-harv}
\bibliography{references}

\end{document}